\documentclass[12pt,showpacs]{revtex4} 
\usepackage{amssymb,epsf}
\usepackage{latexsym}
\usepackage{color}
\begin{document}

\title{\textbf{Phantom phase power-law solution in $f(G)$ gravity}}
\author{A. R. Rastkar}
\affiliation{Department of Physics\\ Azarbaijan University of Tarbiat Moallem, Tabriz, Iran}
\author{M. R. Setare}
\affiliation{Department of Science\\ Payame Noor University, Bijar, Iran}
\author{F. Darabi}
\email{f.darabi@azaruniv.edu} 
\affiliation{Department of Physics\\ Azarbaijan University of Tarbiat Moallem, Tabriz, Iran \\
Corresponding Author}

\date{\today}
\begin{abstract}
\textbf{Abstract:} 
Power-law solutions for $f(G)$ gravity coupled with perfect fluid have been
studied for spatially flat universe. It is shown that despite the matter dominated and accelerating power-law solutions, the power-law solution
exists for an special form of $f(G)$ when this universe enters a Phantom phase. 


\end{abstract}
\maketitle
\newpage
\section{Introduction}

Nowadays it is strongly believed that the universe is experiencing
an accelerated expansion. Recent observations from type Ia
supernovae \cite{SN} in associated with Large Scale Structure
\cite{LSS} and Cosmic Microwave Background anisotropies \cite{CMB}
have provided main evidence for this cosmic acceleration. These
observations also suggest that our universe is spatially flat,
and consists of about $70 \%$ dark energy (DE) with negative
pressure, $30\%$ dust matter (cold dark matter plus baryons), and
negligible radiation. On the other hand the nature of dark energy
is ambiguous. The simplest candidate of dark energy is a
cosmological constant with the equation of state parameter $w=-1$.
However, this scenario suffers from serious problems like a huge
fine tuning and the coincidence problem \cite{4}. Alternative
models of dark energy suggest a dynamical form of dark energy,
which is often realized by one or two scalar fields. In this
respect, dark energy has many dynamical components such as quintessence \cite{5},
K-essence \cite{6}, tachyon \cite{7}, phantom \cite{8}, ghost
condensate and quintom \cite{9}, and so forth.\\
It is known that Einstein's theory of gravity may not describe gravity at very high energies. The simplest alternative to general relativity is
Brans-Dicke scalar-tensor theory \cite{10}. Modified gravity also
provides the natural gravitational alternative for dark energy
\cite{11}. Moreover, thanks to the different roles of gravitational terms relevant at small and at large curvature, the modified gravity presents natural unification of the early-time inflation and late-time acceleration . It may naturally describe the transition from non-phantom phase to phantom one without
necessity to introduce the exotic matter. But among the most
popular modified gravities which may successfully describe the
cosmic speed-up is $f(R)$ gravity. Very simple versions of such
theory like $\frac{1}{R}$ \cite{12} and $\frac{1}{R}+R^2$
\cite{13} may lead to the effective quintessence/phantom
late-time universe (to see solar system constraints on modified
dark energy models refer to \cite{14}). Another theory proposed as
gravitational dark energy is scalar-Gauss-Bonnet gravity \cite{15}
which is closely related with the low-energy string effective action.
In this proposal, the current acceleration of the universe
may be caused by mixture of scalar phantom and (or) potential/stringy effects.
The co–existence of matter dominated and accelerating power law solutions for this theory has already been shown \cite{Goheer0}. In this paper, we
extend these results to show the existence of Phantom phase power law solutions
for an special form of $f(G)$ gravity.

\section{Field equations for  $[R+f(G)+L_m]$ gravity }

As an alternative to the $f(R)$ action considered in Ref.\cite{Goheer}
we consider the following $f(G)$ action which describes Einstein's gravity
coupled with perfect fluid plus a function of the Gauss-Bonnet term \cite{Nojiri1}, \cite{Nojiri2}
\begin{equation}\label{1}
S=\int d^4x \sqrt{-g}\left[\frac{1}{2\kappa^2}R+f(G)+L_m\right]_,
\end{equation}
where $\kappa^2=8\pi G_N$ and the Gauss-Bonnet invariant is defined as follows
\begin{equation}\label{2}
G=R^2-4R_{\mu \nu}R^{\mu \nu}+R_{\mu \nu \lambda \sigma}R^{\mu \nu \lambda \sigma}.
\end{equation}
The field equations are obtained by varying the action with respect to $g_{\mu\nu}$ $$
0=\frac{1}{2k^2}(-R^{\mu\nu}+\frac{1}{2}g^{\mu\nu}R)+T^{\mu\nu}+\frac{1}{2}g^{\mu\nu}f(G)-2f_{G}RR^{\mu\nu}+4f_{G}R^{\mu}_{\rho}R^{\nu\rho}
$$
$$
-2f_{G}R^{\mu\rho\sigma\tau}R^{\nu}_{\rho\sigma\tau}-4f_{G}R^{\mu\rho\sigma\nu}R_{\rho\sigma}+2(\nabla^{\mu}\nabla^{\nu}f_{G})R
-2g^{\mu\nu}(\nabla^{2}f_{G})R-4(\nabla_{\rho}\nabla^{\mu}f_{G})R^{\nu\rho}
$$
\begin {equation}\label{3}
-4(\nabla_{\rho}\nabla^{\nu}f_{G})R^{\mu\rho}+4(\nabla^{2}f_{G})R^{\mu\nu}+4g^{\mu\nu}(\nabla_{\rho}\nabla_{\sigma}f_{G})R^{\rho\sigma}
-4(\nabla_{\rho}\nabla_{\sigma}f_{G})R^{\mu\rho\nu\sigma}\,,
\end{equation}
where $f_G=f'(G)$ and $f_{GG}=f''(G)$. The usual spatially-flat metric of
Friedmann-Robertson-Walker (FRW) universe is chosen in agreement with observations
\begin {equation}\label{4}
ds^{2}=-dt^{2}+a(t)^{2}\sum^{3}_{i=1}(dx^{i})^{2},
\end{equation}
where $a(t)$ is the scale factor as a one-parameter function of the cosmological time $t$. Using this metric in the field equations (\ref{3}) one obtains the first FRW equation 
\begin {equation}\label{5}
-\frac{3}{\kappa^{2}}H^{2}+G 
f_{G}-f(G)-24\dot{G}H^{3}f_{GG}+\rho_{m}=0\,,
\end{equation}
where $.$ denotes derivative with respect to time $t$ and Hubble parameter $H$ is defined by $H=\dot{a}/a$. In the FRW universe, the energy conservation law can be expressed as the standard continuity equation
\begin {equation}\label{6}
\dot{\rho_{m}}+3H(\rho_{m}+p)=\dot{\rho_{m}}+3H(1+w)\rho_{m}=0\ ,
\end{equation}
where $\rho_{m}$ is the matter energy density and $p=w\rho_{m}$ is the equation of state relating pressure $p$ with energy density. From the continuity equation
we obtain
\begin {equation}\label{7}
\rho_{m}(t)=\rho_{0}t^{-3m(1+w)}.
\end{equation}
For the metric (\ref{4}), the Gauss-Bonnet invariant $G$ and the Ricci scalar $R$ may be defined as functions of the Hubble parameter 
\begin {equation}\label{8}
G=24(\dot{H}H^{2}+H^{4}),\quad R=6(\dot{H}+2H^{2}).
\end{equation}

\section{Exact matter dominant and accelerating power-law solutions }

We now assume an exact power-law solution for the field equations
\begin {equation}\label{9}
a(t)=a_0 t^m,
\end{equation}
where $m$ is a positive real number. Using this assumption in (\ref{8}) leads us to the following results
\begin {equation}\label{10}
G=\frac{24}{t^4}m^3(m-1),
\end{equation}
\begin {equation}\label{11}
\dot{G}=-\frac{96}{t^5}m^3(m-1),
\end{equation}
\begin {equation}
R=\frac{6m}{t^2}(2m-1).
\end{equation}
By substituting (\ref{7}), (\ref{10}) and (\ref{11}) into (\ref{5}) we obtain
the Friedmann equation
\begin {equation}\label{12}
\frac{4}{m-1}G^2 
f_{GG}+G f_{G}-f_{G}-\frac{G^{1/2}}{\kappa^2}\left(\frac{3m}{8(m-1)}\right)^{1/2}
+\rho_{0}\left(\frac{G}{24m^3(m-1)}\right)^{\frac{3}{4}m(1+w)}=0_.
\end{equation}
This is a differential equation for the function $f(G)$ in $G$ space. The
general solution of this equation is obtained as
\begin{equation}\label{13}
f \left( G \right)=C_1 G+C_2{G}^{-\frac{1}{4}(m-1)}-\frac{1}{2}\,
\left[\sqrt{\frac{6m(m-1)}{k^4(m+1)^2}}{G}^{\frac{1}{2}} 
+\, A_{mw} {G}^{\frac{3}{4}\,m \left( 1+w \right) } \right]  ,
\end{equation}
where 
\begin {equation}\label{14}
A_{mw}=\frac {8\rho_0\,\left( m-1 \right)  \left(13824 {m}^{9} \left( m-1
 \right)^3  \right) ^{-\frac{1}{4}\,m \left( 1+w \right) }}{4+m\,[3m(w+1)(w+4/3)-15w-19]},
\end{equation}
and $C_{1}, C_{2}$ are arbitrary constants of integration. This solution
is in agreement with the one obtained in \cite{Goheer0} and, as is explained there, we can without any loss of generality assume the constants $C_{1}= C_{2}=0$. Hence, the required form of the function $f(G)$ becomes 
\begin{equation}\label{13'}
f \left( G \right)=-\frac{1}{2}\,
\left[\sqrt{\frac{6m(m-1)}{k^4(m+1)^2}}{G}^{\frac{1}{2}} 
+\, A_{mw} {G}^{\frac{3}{4}\,m \left( 1+w \right) } \right].
\end{equation}
First, we note that a real valued solution for $f(G)$ requires the values
$m\leq0$ or $m\geq1$. While the former leads to a contracting universe and also causes a divergence at $m=-1$ via the first term in the bracket,
the latter is cosmologically desirable for an expanding universe as follows. 

The case $m=1$ leads to $G=0$ and $R=6t^{-2}$ which is the general relativity limit with the power-law
solution
\begin {equation}\label{17}
a(t)=a_0 t,
\end{equation}
and the energy density
\begin {equation}\label{18}
\rho_{m}(t)=\rho_{0}t^{-3(1+w)}.
\end{equation}
According to (\ref{7}), it is easy to see that $m=2/[3(1+w)]$ indicates the reduction to general relativity, so for $m=1$ the equation of state parameter is fixed by $w=-1/3$ which accounts for a negative pressure but not yet an accelerating
universe. 

The case $m>1$, leads to a nonzero real Gauss-Bonnet term $G$ and a positive Ricci scalar $R$. However, in order to avoid divergence in the Gauss-Bonnet term we have to keep $m$ and $w$ away from the values for which $A_{mw}$ diverges according to the following equation
\begin {equation}\label{19}
4+m\,[3m(w+1)(w+4/3)-15w-19]=0.
\end{equation} 
This case with $m>1$ predicts an accelerating universe.
Thus, power-law solutions of the type $a(t)=a_0 t^m$ or $H=\frac{m}{t}$  exist for the actions of the type $[R+f(G)+L_m]$ with $f(G)$ given by (\ref{13})
except for those values of $m$ which satisfy (\ref{19}). 

\section{Exact Phantom phase power-law solution }

One may also study the power-law solutions where the universe enters a phantom
phase leading to a Big Rip singularity. For this case, the general class
of Hubble parameters and cosmological solutions are defined as
\begin {equation}\label{20}
H(t)=\frac{m}{t_s-t},  
\end{equation} 
\begin {equation}\label{20'}
a(t)=a_0 (t_s-t)^{-m},
\end{equation} 
where $t_s$ is the so called ``Rip time'' at future singularity. Again, using
the above solution and repeating the similar calculations we obtain the following results
\begin {equation}\label{21}
\rho_{m}(t)=\rho_{0}t^{3m(1+w)},
\end{equation}
\begin {equation}\label{22}
G=\frac{24}{(t_s-t)^4}m^3(m+1),
\end{equation}
\begin {equation}\label{23}
\dot{G}=\frac{96}{(t_s-t)^5}m^3(m+1),
\end{equation}
\begin {equation}\label{24}
R=\frac{6m}{(t_s-t)^2}(2m+1).
\end{equation}
Substituting (\ref{21}), (\ref{22}) and (\ref{23}) into the first FRW equation (\ref{5}) we obtain 
\begin {equation}\label{24'}
-\frac{4}{m+1}G^2 
f_{GG}+G f_{G}-f_{G}-\frac{G^{1/2}}{\kappa^2}\left(\frac{3m}{8(m+1)}\right)^{1/2}
+\rho_{0}\left(\frac{G}{24m^3(m+1)}\right)^{-\frac{3}{4}m(1+w)}=0_.
\end{equation}
This equation is easily recovered by the map $m \rightarrow -m$ in the previous
equation (\ref{12}). Therefore its solution is obtained by using the same map in
(\ref{13}) as
\begin{equation}\label{25}
f \left( G \right)=C_1 G+C_2{G}^{\frac{1}{4}(m+1)}-\frac{1}{2}\,
\left[\sqrt{\frac{6m(m+1)}{k^4(m-1)^2}}{G}^{\frac{1}{2}} 
+\, A_{mw} {G}^{-\frac{3}{4}\,m \left( 1+w \right) } \right]  ,
\end{equation}
where 
\begin {equation}\label{26}
A_{mw}=-\frac {8\rho_0\,\left( m+1 \right)  \left(13824 {m}^{9} \left( m+1
 \right)^3  \right) ^{\frac{1}{4}\,m \left( 1+w \right) }}{4+m\,[3m(w+1)(w+4/3)+15w+19]}.
\end{equation} 
Similar to the solutions in the previous section, we assume $C_{1}= C_{2}=0$.
Then, the required form of the function $f(G)$ becomes  
\begin{equation}\label{25'}
f \left( G \right)=-\frac{1}{2}\,
\left[\sqrt{\frac{6m(m+1)}{k^4(m-1)^2}}{G}^{\frac{1}{2}} 
+\, A_{mw} {G}^{-\frac{3}{4}\,m \left( 1+w \right) } \right].
\end{equation}
Actually, $m>0$ leads to a real valued function $f(G)$ according to (\ref{25}).
However, demanding a Big Rip during the phantom phase, as the cosmic time $t$ approaches $t_s$, requires $m\geq1$ in (\ref{20'}). However, $m=1$ causes a divergence in $f(G)$ through the first term of (\ref{25}) in the bracket. Moreover,  the Gauss-Bonnet term diverges through $A_{mw}$ for those values of $m$ for which the following equation is satisfied
\begin {equation}\label{27}
4+m\,[3m(w+1)(w+4/3)+15w+19]=0.
\end{equation} 
Therefore, power-law solutions in the phantom phase of the type $a(t)=a_0 (t_s-t)^{-m}$ or $H(t)=\frac{m}{t_s-t}$ exist for the actions of the type $[R+f(G)+L_m]$ with $f(G)$ given by (\ref{25})
except for those values of $m$ which satisfy (\ref{27}). 

\section{\large The stability issue}

The stability issue of a large class of modified gravitational models has
been discussed with particular emphasis to de Sitter solutions \cite{Cogn2,Cogn3,
Cogn4,Cogn5,Capo2,Eliz1},\cite{Fara1},\cite{Fara2}\cite{NojiIII}, \cite{Noji11111}.
In the present modified gravity, namely $[R+f(G)]$, the stability
issue leads to the following conditions \cite{Cogn4}
\begin{equation}\label{28}
G_0f^{'}_0-f_0=6H_0^2,
\end{equation}
\begin {equation}\label{29}
1<\frac{9}{R_0^3f^{''}_0},
\end{equation}
where the critical points are defined by
\begin{equation}\label{30}
R_0=12H_0^2,\:\:  G_0=24H_0^4,
\end{equation}
and $f_0$, $f^{'}_0=(df/dG)_0$, $f^{''}_0=(d^2f/dG^2)_0$, $H_0$ are suitable constants corresponding to the de Sitter solutions. In the case of Phantom phase power law solution, namely (\ref{25'}), the first condition reads as
\begin{eqnarray}\label{31}
\frac{1}{2}\sqrt{\frac{6m(m+1)}{k^4(m-1)^2}}&+&\left(1-\frac{3}{2}m(1+\omega)\right)A_{mw}
\\ \nonumber
&\times&{G_0}^{-\frac{3}{4}\,m (1+w)-\frac{1}{2} }=\sqrt{6},
\end{eqnarray}
which implies that
\begin{eqnarray}\label{32}
\left(1-\frac{3}{2}m(1+\omega)\right)A_{mw}>0.
\end{eqnarray}
By using (\ref{32}), the second condition (\ref{29}) reads as
\begin{eqnarray}\label{33}
&&(3/2)^{3/2}\sqrt{\frac{6m(m+1)}{k^4(m-1)^2}}+\left(-\frac{3}{4}m(1+\omega)\right)\\
\nonumber
&-&
\left(\frac{3}{4}m(1+\omega)+1\right)\left(1-\frac{3}{2}m(1+\omega)\right)^{-1}\\
\nonumber
&\times&\left[\sqrt{6}-\frac{1}{2}\sqrt{\frac{6m(m+1)}{k^4(m-1)^2}}\right]<9.
\end{eqnarray}
Then, the model is stable around de Sitter solution if the arbitrary parameters also satisfy both the conditions (\ref{32}) and (\ref{33}).

\section{Conclusion}

In the present paper we have considered an $f(G)$ action which
describes Einstein's gravity plus a function of the Gauss-Bonnet
term. Then, by considering an exact power-law solution for the
field equations we have obtained the Friedmann equation in
spatially flat universe. The Friedmann equation appears as a
differential equation for function $f(G)$. We could obtain the
solution of this equation and show that our model with this solution for
$f(G)$ has power-law solution of the type $a(t)=a_0 t^m$ expect for those values of $m$ for which $f(G)$ diverges. These solutions are in agreement
with those obtained in \cite{Goheer0}. 
We have also studied the power-law solutions when the
universe enters a Phantom phase. By considering such power-law
solution for the field equations we have obtained the corresponding Friedmann
equation. The solution $f(G)$ of this differential equation is obtained and it is shown that the power-law solution in the phantom phase of the type $a(t)=a_0(t_s-t)^{-m}$ exists for this $f(G)$ except for those values of $m$ for which this function diverges.

\section*{Acknowledgment}
This work  has been supported by the Research office of Azarbaijan
University of Tarbiat Moallem, Tabriz, Iran.

\end{document}